\documentclass[prb,showpacs,aps]{revtex4}
\usepackage{graphicx}
\usepackage{dcolumn}
\begin{document}
\title{Electron Exchange Coupling for Single Donor Solid-State Spin Qubits}
\author{C.J. Wellard$^a$, L.C.L. Hollenberg$^a$, F. Parisoli$^{a,b}$, L. Kettle$^{c}$, H.-S. Goan$^{d}$, J.A.L. McIntosh$^a$ and D.N. Jamieson$^a$}
\affiliation{Centre for Quantum Computer Technology,\\
$^a$ School of Physics, University of
Melbourne, Victoria 3010, AUSTRALIA. \\
$^b$ Department of Physics, University of Bologna, Bologna 40126, ITALY\\
$^c$ University of Queensland, QLD 4072, AUSTRALIA \\
$^d$ University of New South Wales, Sydney NSW 2052, AUSTRALIA.
}
\date{\today}
\begin{abstract}
Inter-valley interference between degenerate conduction band
minima has been shown to lead to oscillations in the exchange
energy between neighbouring phosphorus donor electron states in silicon
\cite{Koiller02,Koiller02A}. These same effects lead to an extreme
sensitivity of the exchange energy on the relative orientation of
the donor atoms, an issue of crucial importance in the construction
silicon-based spin quantum computers. In this article we calculate
the donor electron exchange coupling as a function of donor
position incorporating the full Bloch structure of the
Kohn-Luttinger electron wavefunctions. 
It is found that due to the rapidly oscillating nature of the terms they 
produce, the periodic part of the Bloch functions can be safely ignored in 
the Heitler-London integrals as was done by Koiller {\it et al} \cite{Koiller02,Koiller02A}, significantly reducing the complexity of calculations.
 We address issues of fabrication and
calculate the expected exchange coupling between neighbouring
donors that have been implanted into the silicon substrate using
an 15keV ion beam in the so-called 'top down' fabrication scheme
for a Kane solid-state quantum computer. In addition we calculate
the exchange coupling as a function of the voltage bias on control
gates used to manipulate the electron wavefunctions and implement
quantum logic operations in the Kane proposal, and find that these gate biases 
can be used to both increase and decrease the magnitude of the exchange
coupling between neighbouring donor electrons. The zero-bias results reconfirm those previously obtained by Koiller {\it et al} \cite{Koiller02A}.
\end{abstract}
\pacs{ 03.67.Lx, 71.55.Cn, 85.30.De}
\maketitle
\section{introduction}
Solid state systems have emerged as a promising candidate for the
construction of a large scale quantum computer (QC). In particular spin based
architectures take advantage of the relatively long spin dephasing
times of donor electrons or nuclei in silicon. Single qubit
operations are performed by tuning the spin Zeeman energy to be
resonant with an oscillating field which drives the transition
while neighbouring qubits are coupled via the electron exchange
interaction, whether it be directly in the case of electron-spin
proposals \cite{Vrijen00}, or indirectly in the case of nuclear
spin quantum computers.
\par
\begin{figure}
\rotatebox{0}{\resizebox{8cm}{!}{\includegraphics{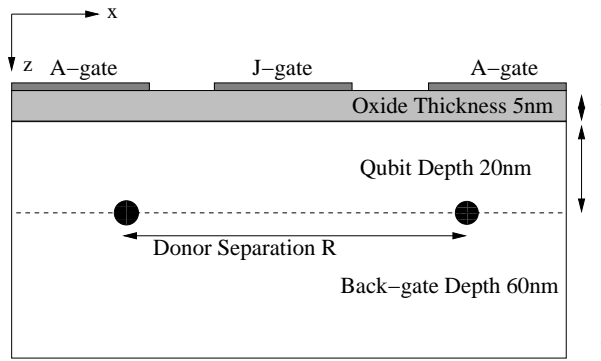}}}
\caption{\label{fig:kanepic}The Kane architecture based on buried
phosphorus dopants in a silicon substrate}
\end{figure}

 In this work we concentrate
on the Kane\cite{Kane98} concept of single phosphorus donor
nuclear spin qubits embedded in a silicon substrate, which is a
leading candidate in the search for a scalable QC architecture.
The Kane architecture (shown in Fig.~\ref{fig:kanepic}.) calls for
the placement of phosphorus donors at subsitutional (face-centred cubic (fcc)) 
sites in the host silicon matrix, with inter-donor
spacings of order 200\AA. Quantum logic operations on the
nuclear-spin qubits are implemented through coherent control of
the donor electron wavefunctions which are coupled to the donor
nuclei through the contact hyperfine interaction. This control of
the electron wavefunctions is achieved through application of
voltage biases to control gates placed on the substrate surface
above (A-gate), and between (J-gate) the phosphorus donors, which
create electrostatic potentials within the device thus altering
the form of the electron wavefunctions.
\par In a recent paper, Koiller, Hu and Das
Sarma (KHD) \cite{Koiller02} presented theoretical evidence for
oscillations in the electron-mediated exchange coupling as a
function of inter-donor separation, and a strong dependence of
this coupling on the relative orientation of the two donors with
respect the structure of the silicon substrate. They calculate the
exchange coupling using an approximate version of the
Heitler-London formalism and by essentially ignoring the periodic
part of the Bloch wavefunctions in the expression for the donor
electron wavefunctions. In this article we eliminate both these
approximations, by calculating the exchange coupling in the full
Heitler-London formalism and by including the full Bloch structure
of the donor electron wavefunction. We show that while the first
approximation breaks down for small donor separations, as
discussed by KHD themselves \cite{Koiller02A}, the second
approximation, that of ignoring the periodic part of the Bloch
functions, is in excellent agreement with the full calculations
regardless of the donor orientation.
\par This paper is organised as follows. In section~\ref{section:dwf} we review the calculation of the
valence electron wavefunctions for phosphorus donors in silicon,
in the Kohn-Luttinger effective mass formalism. In
section~\ref{section:hlf} we discuss the Heitler-London formalism
used to calculate the exchange coupling between neighbouring donor
electrons, and we show that the terms that arise due to the
periodic part of the Bloch functions oscillate sufficiently
rapidly so as to average to zero over the range of the integrals.
Section~\ref{section:fab}  contains a discussion of some of the
fabricational issues that arise from the strong position
dependence of the exchange energy on the production of a
phosphorus donor based solid state quantum computer.
\par Section~\ref{section:vdep} is devoted to a calculation of the
dependence of the exchange coupling between neighbouring donor
electrons on voltage biases applied to the control J-gate, used to
tune the inter-qubit coupling. By extending the conventional
effective mass formalism we construct a basis of non-isotropic
hydrogen like envelope functions in which we expand the
donor-electron wavefunction in the presence of the electric
potentials. The potential created inside the device due to the
J-gate bias is calculated using a commercial software package
specifically designed for the modelling of semiconductor devices.
Donor wavefunctions are obtained by direct diagonalisation, and
the Heitler-London formalism is used to determine the exchange
coupling for various gate biases and donor separations oriented in
the silicon [100] direction.
\section{The donor electron wave function}
\label{section:dwf}
Although the qubits of the Kane quantum computer are encoded by
the nuclear spins, it is the donor electron which mediates both
single and coupled gate operations. Therefore, the crucial element
in the quantum description of the device is the donor electron
wave function. Initial attempts to describe the operation of the
device, particularly in response to external time dependent gate
potentials which complicate the situation considerably, have
focused on effective hydrogenic-type approximations for the donor
electron wave function\cite{Wellard02,Kettle03b}, or the use of 
simplified potentials \cite{Fang02}. These calculations provide some useful
estimates, however more detailed calculations are required, 
using realistic models of both the donor 
electron wavefunction and the potentials induced inside the device by the 
application of control gate biases.
\par In going beyond the effective hydrogenic treatment of the
ground-state electron wave function for phosphorus donors in
silicon it is recognised that the underlying crystal structure of
the silicon must have an effect. The wavefunction is thus
expanded in the basis of the Bloch functions for silicon,
\cite{Kohn55,Kohn57}
\begin{equation}
\Psi({\bf r}) = \int F({\bf k}) \phi_{\bf k}({\bf r}) d{\bf k}.
\label{equation:donor_wf1} 
\end{equation}
The coefficients $F({\bf k})$ are obtained by substituting the
above form into the Schr\"odinger equation, with the Hamiltonian
$H = H_0 - U(r)$, where $H_0$ is the Hamiltonian for the pure
silicon crystal and $U(r)$ is the donor-potential for phosphorus.
The Bloch functions can be written in the form $\phi_{\bf k}({\bf
r}) = {\rm e}^{i{\bf k}.{\bf r}} u_{\bf k} ({\bf r})$, where $u_{\bf k}
({\bf r})$ is a function that shares the periodicity of the
lattice, and can be expanded in a basis of plane waves with
wave-vectors equal to the reciprocal lattice vectors for silicon
${\bf G}$, $u_{\bf k}({\bf r}) = \sum_{\bf G} A_{\bf k}({\bf
G}) {\rm e}^{i {\bf G}.{\bf r}}$. The result is that the
Schr\"odinger equation can be written as
\begin{eqnarray}
E F({\bf k}) &=& E^0_{\bf k} F({\bf k}) + \sum_{{\bf G},{\bf G}'}
\int A_{{\bf k}',{\bf G}'}^* A_{{\bf k},{\bf G}}  U({\bf r}) {\rm
e}^{i({\bf k}-{\bf k}').{\bf r}}{\rm e}^{i({\bf G} -{\bf G'}).{\bf
r}}  F({\bf k}') d{\bf r} d{\bf k}'\nonumber\\ &=& E^0_{\bf k}
F({\bf k}) + \sum_{{\bf G},{\bf G}'} \int A_{{\bf k}',{\bf G}'}^*
A_{{\bf k},{\bf G}}  {\tilde U}({\bf k}+{\bf G} - {\bf k}' -{\bf
G}')  F({\bf k}') d{\bf k}', \label{equation:schroedinger}
\end{eqnarray}
where ${\tilde U}({\bf k}) = \int U({\bf r}) {\rm e}^{-i {\bf
r}.{\bf k}} d{\bf r}$ is the Fourier transform of the impurity
potential. The $E^0_{\bf k}$  are the eigenenergies of the Bloch
functions $\phi_{\bf k}({\bf r})$, for the pure silicon lattice.
We make the approximation that due to the increased energies of
the higher bands, only conduction band states contribute to the
impurity wavefunction. Further, in silicon there are six degenerate
conduction band minima, located along the $\langle 100 \rangle$
directions in ${\bf k}$-space, 85\% of the way between the centre
($\Gamma$ point), and the zone boundary ($X$ point). Because of
the reduced energies in these regions the envelope functions can
be expressed as a sum of functions localised around each of the
conduction band minima $F({\bf k}) = \sum_\mu F_\mu({\bf k})$.
\par In the effective-mass treatment \cite{Kohn55,Kohn57,Pantelides78}
the Bloch energies are expanded to second order around the
conduction band minima $E^0_{\bf k} \approx \frac{\hbar^2}{2}
\left(\frac{{\bf k}_\perp^2}{m_\perp}
+\frac{k_\parallel^2}{m_\parallel} \right)$, where ${\bf k_\perp}$ is
the component of ${\bf k} - {\bf k_\mu}$ perpendicular to ${\bf
k}_\mu$ and $k_\parallel$ is the parallel component. The
$m_\perp,m_\parallel$ are effective masses and the inequality of
these two values reflects the anisotropy of the conduction band
minima. An additional approximation is made whereby only the terms
with ${\bf G} = {\bf G}'$ in the potential term of the
Schr\"odinger equation are included. The assumption is that
$U({\bf k}-{\bf k}'+{\bf G} -{\bf G}') << U({\bf k} - {\bf k}')$
for ${\bf G} \neq {\bf G}'$. This approximation is well satisfied
for a Coulombic potential $U({\bf k}) = 1/(\kappa \pi^2 |{\bf
k}|^2)$, with $\kappa = 11.9$ the dielectric constant for silicon,
and with the reciprocal lattice vectors of magnitude $|{\bf G}| =
n 2 \pi/d$, where $n$ is an integer and $d = 5.43$\AA \ is the
lattice spacing for silicon. Another way of viewing the
approximation is that the terms ${\rm e}^{i({\bf G} -{\bf
G}').{\bf r}}$ in the first line of Eq.~\ref{equation:schroedinger}
oscillate on a scale sufficiently short compared to variations in
$U({\bf r})$, that they average the integrand to zero. We will
show in the next section that the same approximation allows us to
ignore the periodic part of the Bloch functions in the donor
electron wavefunctions when performing Heitler-London integrations.
\par One further approximation is necessary,
$A_{{\bf k'},{\bf G}} \approx A_{{\bf k},{\bf G}}$, which coupled
with the relation $\sum_{\bf G} |A_{{\bf k},{\bf G}}|^2 =1$ gives
the effective mass Schr\"odinger equation
\begin{equation}
\frac{\hbar^2}{2} \left(\frac{{\bf k}_\perp^2}{m_\perp}
+\frac{k_\parallel^2}{m_\parallel}\right)\sum_\mu F_\mu({\bf k}) + \int
{\tilde U}({\bf k} -{\bf k}') \sum_\mu F_\mu({\bf k}') d{\bf k}' = E
\sum_\mu F_\mu({\bf k}).
\end{equation}
In the standard effective mass treatment the so-called
valley-orbit terms, which couple the envelope functions at different
conduction band minima are ignored, and we are left with six
independent equations, one for each minimum. With a Coulombic impurity
potential the solutions are non-isotropic hydrogenic wavefunctions
of the form
\begin{equation}
F_{\pm z}({\bf r}) = \frac{ \exp \left[ -\sqrt{(x^2+y^2)/a_{\perp}
+ z^2/a_{\parallel} } \,\,\right]}{\sqrt{6 \pi a_{\perp}^2
a_{\parallel}}}, \label{equation:envelope}
\end{equation}
where $F_\mu({\bf r}) = \int F_\mu({\bf k}-{\bf k}_\mu) {\rm e}^{i
{\bf k}.{\bf r}} d{\bf k}$, and we have used the
example of the envelope function localised around the $z$-minima
of the conduction band. The Kohn-Luttinger\cite{Kohn55,Kohn57}
form for the electron wavefunction of a donor situated at any
position ${\bf R}$ is then given by
\begin{equation}
\Psi({\bf r-R}) = \sum_{\mu} F_\mu({\bf r-R}) {\rm e}^{\bf k_\mu
.(r-R) } u_\mu ({\bf r}), \label{equation:donor_wf},
\end{equation}
where the periodic part of the Bloch function is indepedant of the position of the substitutional impurity.
The values $a_{\perp} = 25.09 {\rm \AA},a_{\parallel} = 14.43 {\rm
\AA}$, are the effective Bohr radii, and are determined
variationally \cite{Kohn55,Koiller02}.
\par It is well known that the valley orbit coupling does contribute to the
the energy of the state \cite{Fritzsche62}, particularly if the
donor potential is not Coulombic as is the case in the immediate
vicinity of the donor nucleus. The consequence of this is that the
donor electron binding energy given by this wave function is $E =
28.95$meV, significantly lower than the experimental
value of $E=45.5$meV \cite{Aggarwal65,Faulkner69}. This
discrepancy is thought to arise from the deviation of the donor
potential from a purely Coulombic potential as well as the effect
of a non-static dielectric constant in silicon \cite{Pant74}. It
is expected however that at distances of more than approximately
an effective Bohr radius from the donor nucleus, the donor
potential should be approximately Coulombic and so Eq.~\ref{equation:donor_wf} will provide
 a good description of the true donor electron in this region\cite{Kohn55}. 
Thus the Kohn-Luttinger form of the donor-electron wavefunction
should be adequate for the purposes of calculating exchange energies for donor 
separations in the range considered in this article.
\par
\begin{figure}
\rotatebox{90}{\resizebox{5cm}{!}{\includegraphics{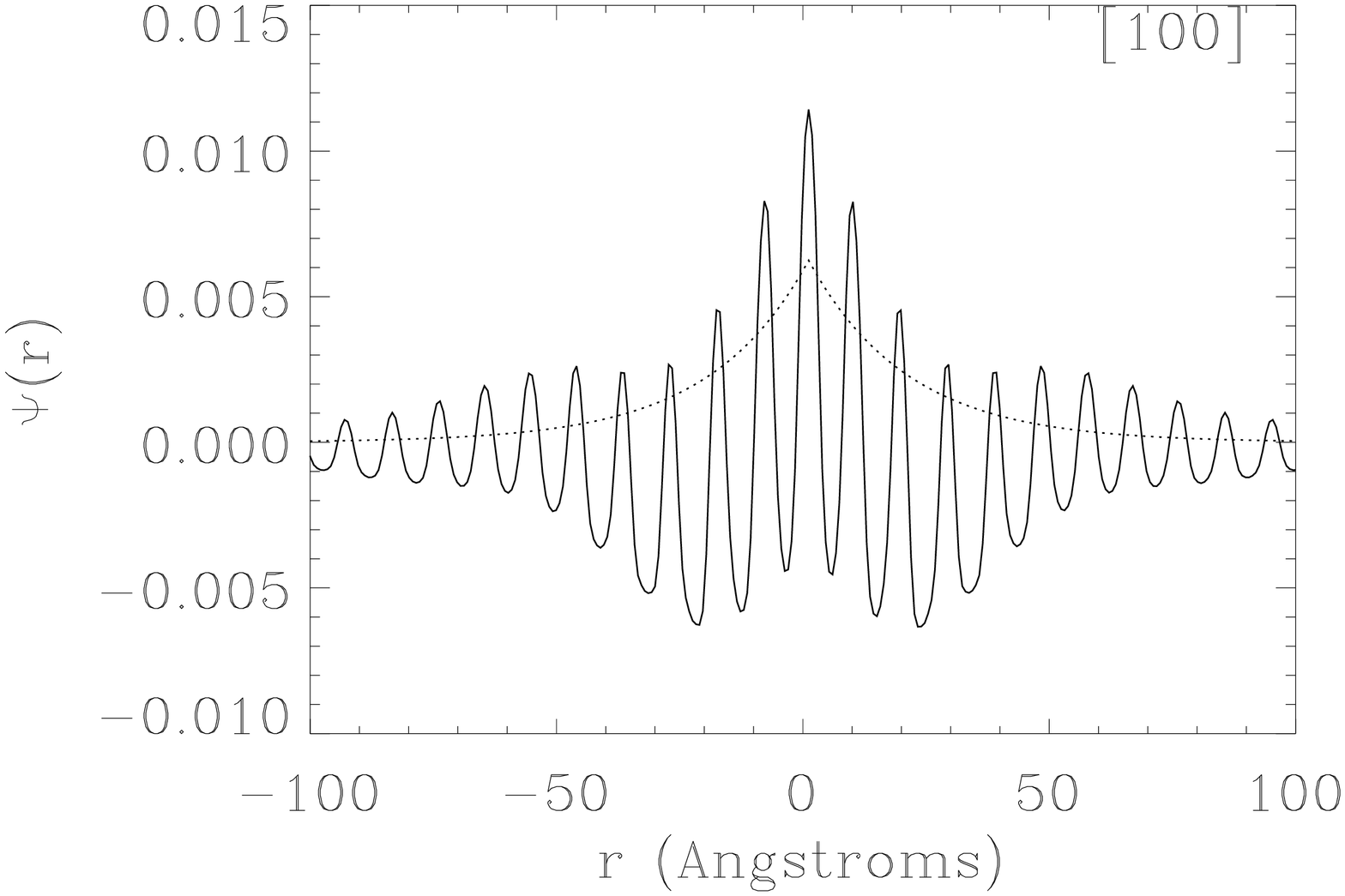}}}
\rotatebox{90}{\resizebox{5cm}{!}{\includegraphics{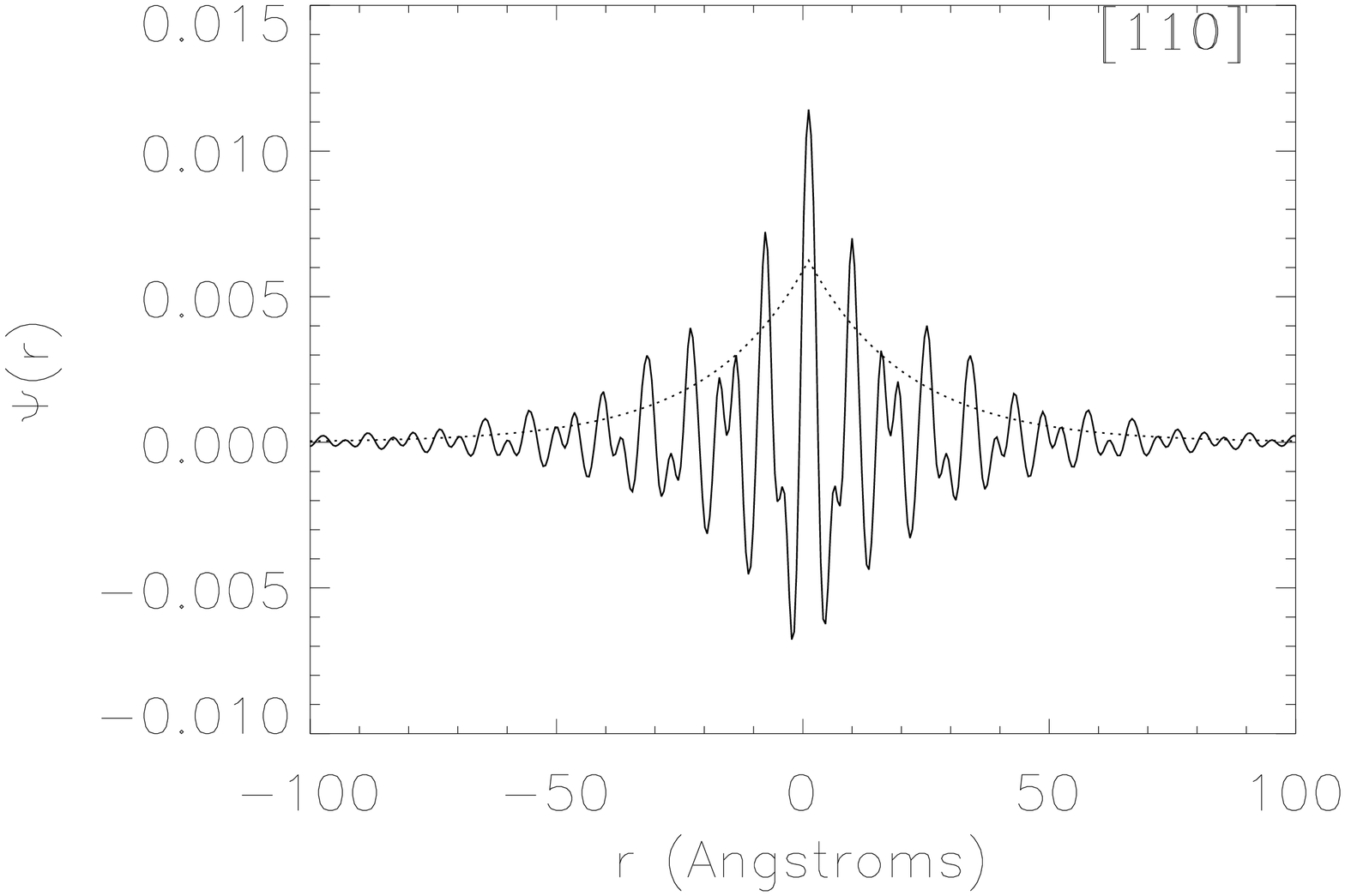}}}
\rotatebox{90}{\resizebox{5cm}{!}{\includegraphics{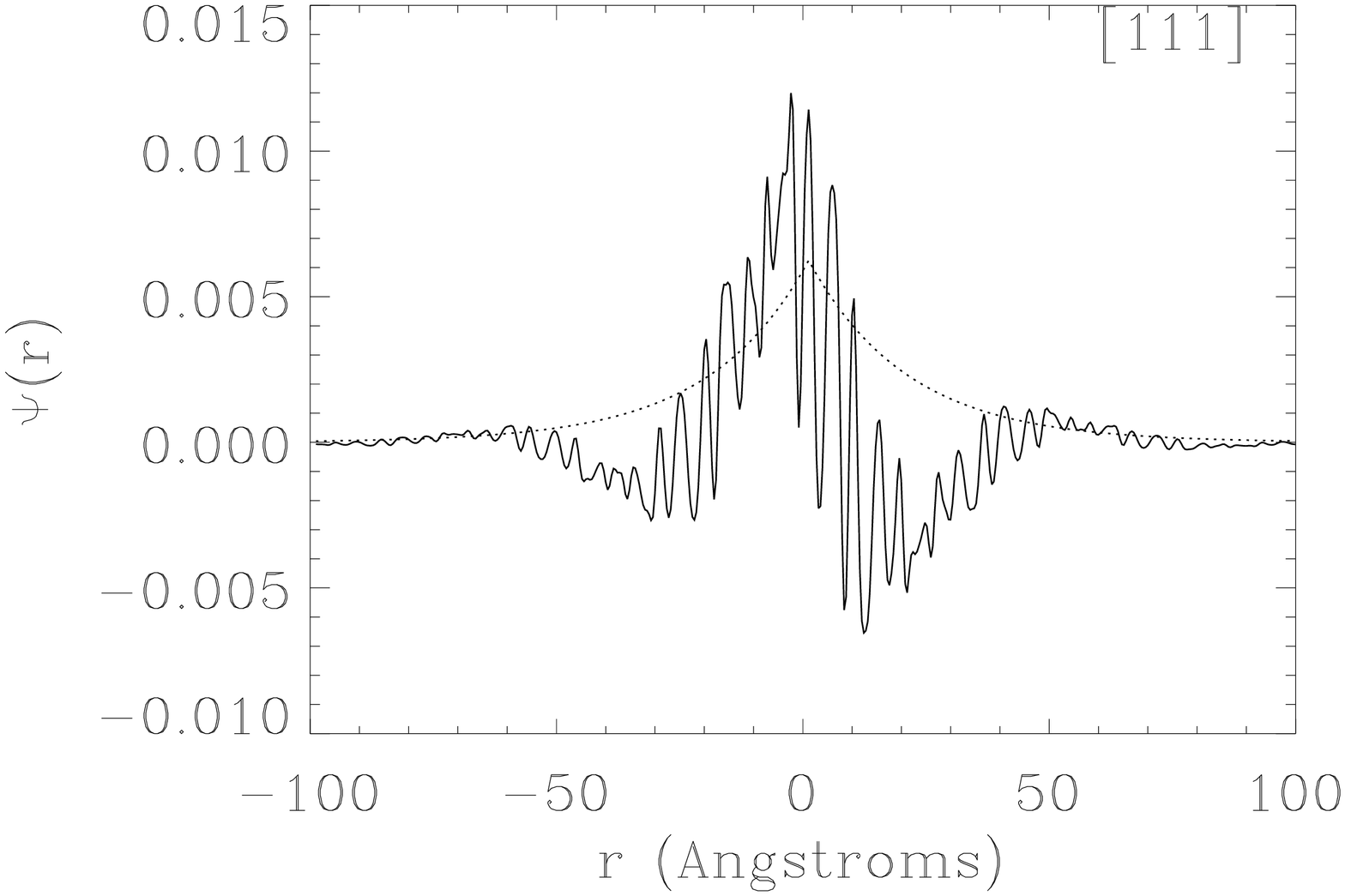}}}
\caption{\label{fig:wf}The solid line shows Kohn-Luttinger wave
function for a phosphorus donor electron in silicon, plotted along
directions of high symmetry within the crystal. The dotted line shows an 
isotropic $1s$ hydrogenic wave function, with an Bohr radius of $20.13$ \AA.}
\end{figure}

 A plot of the donor electron wave function along three directions
of high symmetry, calculated using effective Bohr radii as
determined by Koiller {\it et al} \cite{Koiller02}, is shown in
Fig.~\ref{fig:wf} for a donor placed at a substitutional lattice site. The
coefficients $A_{{\bf k}_\mu,{\bf G}}$, were calculated using the
simple local empirical pseudopotential method as outlined in
reference \cite{Cohen88}, and a basis of 125 states. This method
accurately reproduces the electronic band structure for silicon, particularly
in the region of interest for this calculation, the lowest
conduction bands, as obtained using more complicated non-local
pseudo-potential techniques.
The donor electron wave functions obtained in this manner clearly
display oscillations produced by the interference between the
Bloch functions located at the different conduction band minima.
The wavefunctions are real, and in the [111] direction slightly
asymmetric, the asymmetry being a consequence of the presence of the
second sub-lattice. In the [111] orientation the silicon atoms are
nor evenly spaced, and so the neighbouring silicon atom on one
side of the phosphorus donor is much closer than that on the other
side. Superimposed over the donor wave function is an isotropic
$1s$ hydrogenic envelope with a Bohr radius of $20.13$ \AA \ ,
reflecting the effect of the superposition of the non-isotropic
envelope functions $F_{\mu}$.
\section{The Heitler London formalism}
\label{section:hlf}
The two electron Hamiltonian for a system of two donors separated by a vector ${\bf R}$, in
effective Rydberg units is:
\begin{eqnarray}
H &=& - {\hbar\over {2 m^2}}{\bf \nabla}_1^2-{\hbar\over {2
m^2}}{\bf \nabla}_2^2 - {e^2\over 4 \pi \epsilon {\bf r}_1} -
{e^2\over 4 \pi \epsilon|{\bf R}-{\bf r}_1|} \nonumber\\
&-& {e^2\over 4 \pi \epsilon{\bf r}_2} - {e^2\over 4 \pi
\epsilon|{\bf R}-{\bf r}_2|} + {e^2\over 4 \pi \epsilon|{\bf
r}_1-{\bf r}_2|} \nonumber\\ &+& {e^2\over 4 \pi \epsilon{\bf R}} +
V_{Si}({\bf r}_1,{\bf r}_2).
\end{eqnarray}
In the standard Heitler-London approximation\cite{Slater63} one
assumes that the lowest energy two electron wavefunction of the
two-donor system is simply a correctly symmetrised superposition of single electron
wavefunctions centred around each of the two donors
\begin{equation}
\Psi_\pm({\bf r}_1,{\bf r}_2)= \frac{1}{\sqrt 2} \left(\Psi({\bf
r}_1 -{\bf R}/2 )\Psi({\bf r}_2 +{\bf R}/2 ) \pm \Psi({\bf r}_1
+{\bf R}/2 )\Psi({\bf r}_2 -{\bf R}/2 )\right),
\end{equation}
where the two donors are located at positions $\pm {\bf R}/2$.
This approximation is asymptotically exact and should hold for
separations greater than the effective Bohr radii, $|{\bf R}| >>
a_\perp,a_\parallel$. The anti-symmetry of the fermion
wavefunction then tells us that the exchange splitting, the
difference in energies between the spin singlet and triplet spin
states, is simply equal to the difference in energy between the
states $\Psi_\pm$, that is $E_{\rm triplet} - E_{\rm singlet} =
E_--E_+$.
\par In this article we present our results in terms of
the exchange, or J coupling, in the exchange term of the
effective spin Hamiltonian, $J {\vec \sigma}_1^e .{\vec \sigma}_2^e$, which is common to the
quantum computing literature. We make this decision despite the
observation that it is the exchange splitting, the energy
difference between the single and triplet states, that is most
commonly presented in the solid-state literature. These quantities
are related by the expression $J = (E_{\rm triplet} - E_{\rm
singlet})/4$.
\par In the Heitler-London formalism the exchange
coupling can  be expressed in terms of matrix elements of the
Hamiltonian one can rewrite this as:
\begin{equation}
J({\bf R}) = {1\over 2}S({\bf R})^2(I_1({\bf R}) - I_2({\bf
R}))/(1-S({\bf R})^4),
\end{equation}
where the overlap integrals are given explicitly by:
\begin{eqnarray}
I_1({\bf R}) &=& \int d^3 {\bf r}_1\,d^3 {\bf r}_2\, \Psi^*({\bf
r}_1+{\bf R}/2) \Psi^*({\bf r}_2 - {\bf R}/2) H \,\Psi({\bf
r}_1+{\bf R}/2) \Psi({\bf r}_2 - {\bf R}/2),\nonumber\\ I_2({\bf
R}) &=& \int d^3 {\bf r}_1\,d^3 {\bf r}_2\, \Psi^*({\bf r}_1+{\bf
R}/2 ) \Psi^*({\bf r}_2 - {\bf R}/2) H \,\Psi({\bf r}_2+{\bf R}/2)
\Psi({\bf r}_1 - {\bf R}/2),\nonumber\\ S({\bf R}) &=& \int d^3
{\bf r} \, \Psi({\bf r}+{\bf R}/2)^*\Psi({\bf r} - {\bf R}/2).
\end{eqnarray}
Computation of these expressions using the Kohn-Luttinger wave
function without approximation is rather tedious, given the large number reciprocal lattice
vectors it is necessary to sum over to evaluate the periodic part of the  Bloch functions
for each of the six Bloch states in the donor electron wavefunction.
We have used an adaptive Monte-Carlo quadrature
program to evaluate all overlap integrals, taking due care to
obtain reasonable precision.
\par
\begin{figure}
\rotatebox{90}{\resizebox{5cm}{!}{\includegraphics{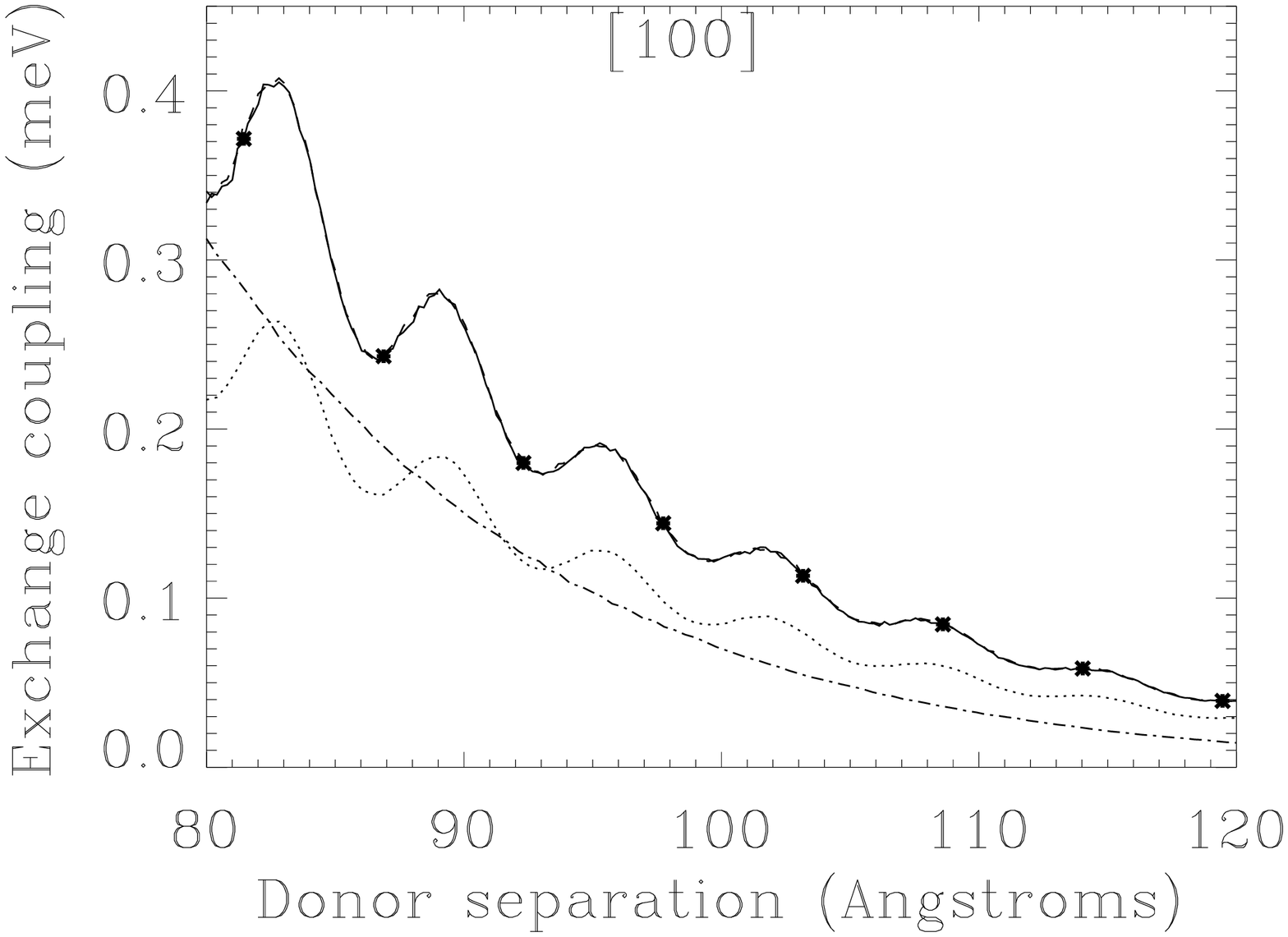}}}
\rotatebox{90}{\resizebox{5cm}{!}{\includegraphics{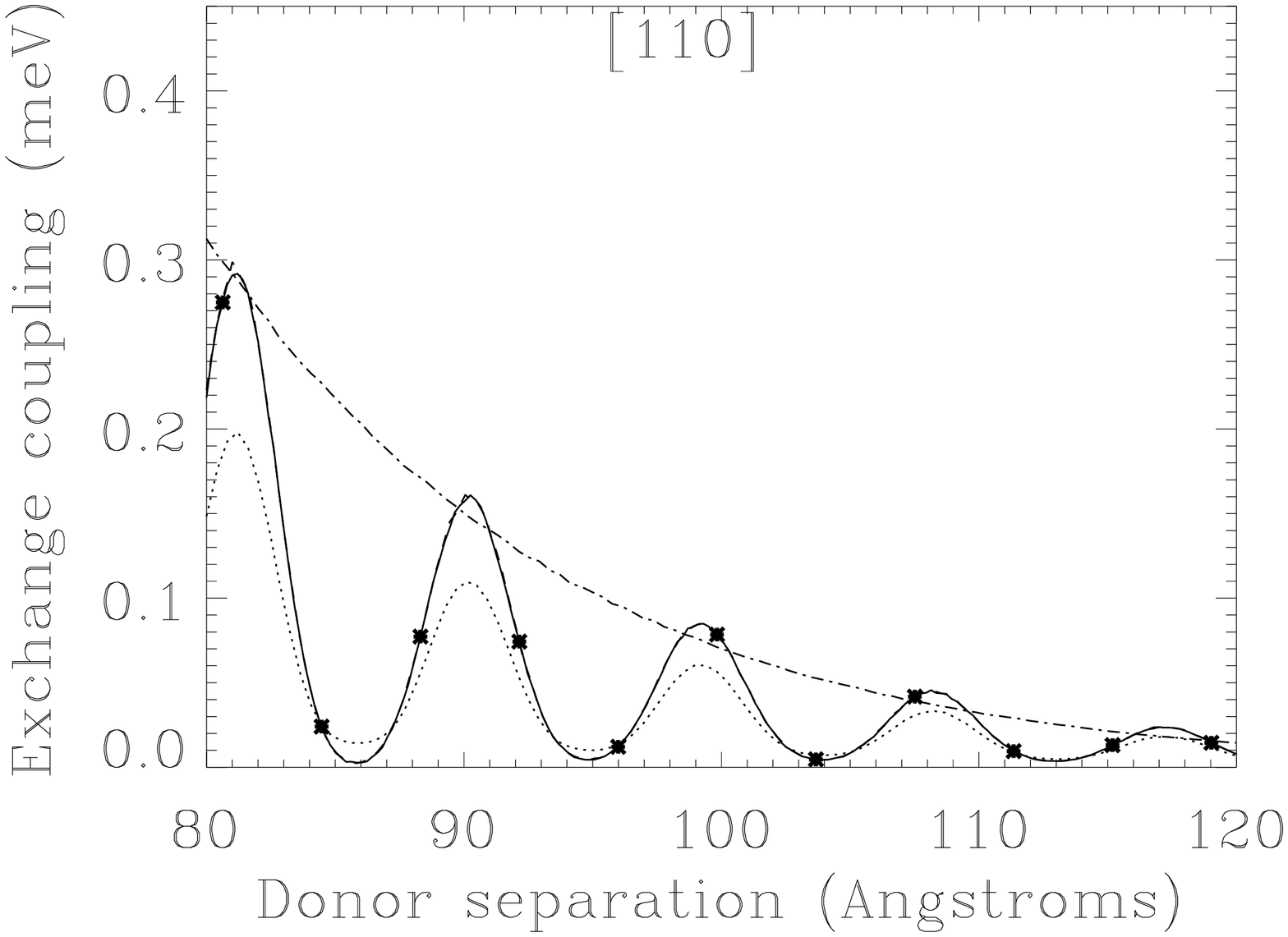}}}
\rotatebox{90}{\resizebox{5cm}{!}{\includegraphics{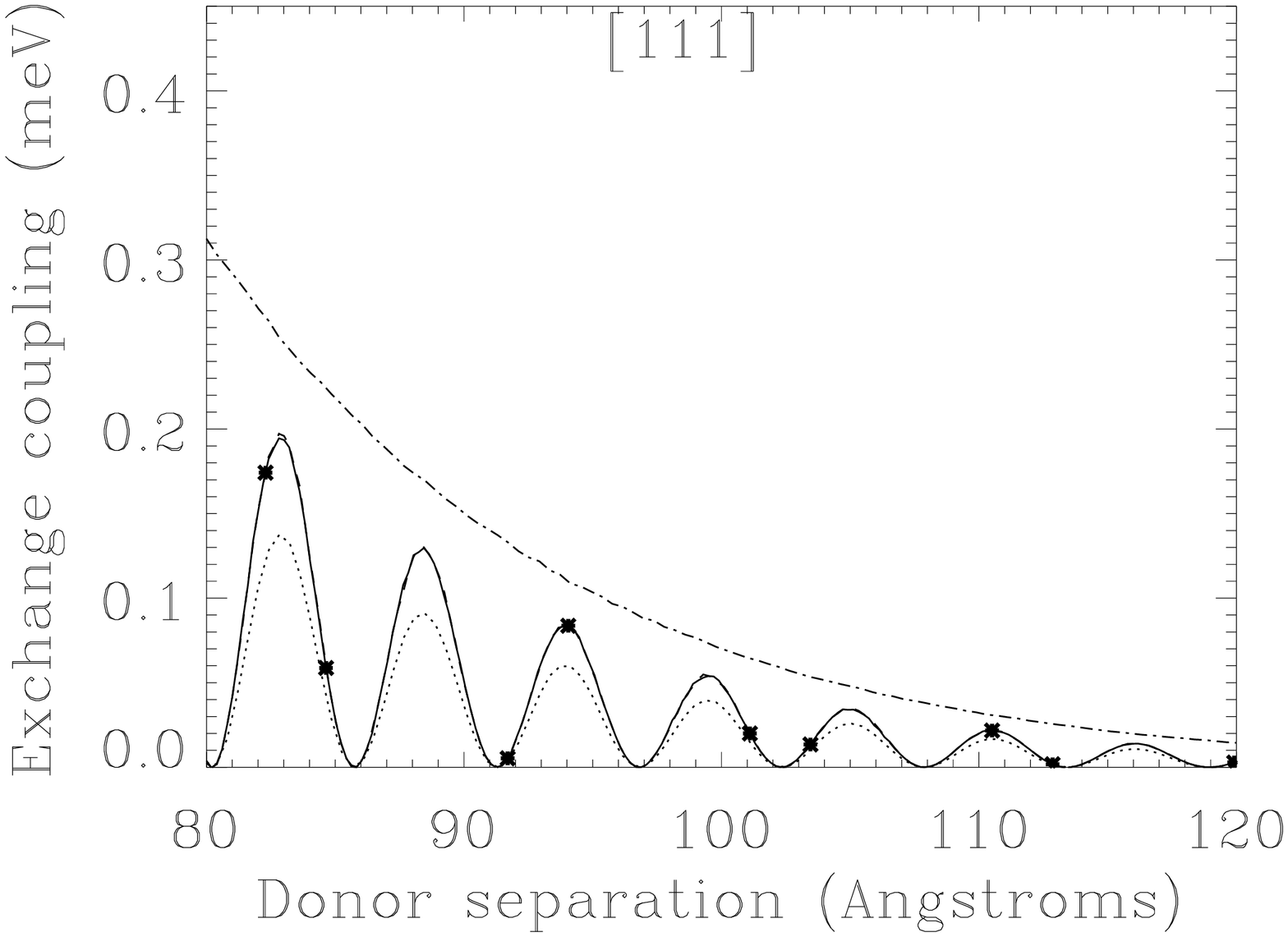}}}
\caption{\label{fig:J} Calculated exchange couplings as a function
of donor separations along three high symmetry directions. In each
case we plot with a solid line the results obtained when the
periodic part of the Bloch function, $u(r)$, is included, a dashed
line indicates the results obtained with $u(r)=1$. In each plot 
these lines cannot be separately resolved. The dotted
line gives the results obtained when using the approximations of
KHD and  the dotted-dashed line gives the exchange coupling
calculated using a 1s hydrogenic orbital with an effective Bohr
radius of 20.13\AA. The $*$ denote the positions of lattice
sites.}
\end{figure}

 In Fig.~\ref{fig:J} we plot the
exchange energy as a function of donor separation in each of the
major lattice directions. For comparison the result using an
isotropic hydrogenic wave function is also given.
We compare our results with those obtained using the method of KHD in their original paper \cite{Koiller02}, who used two major approximations in order to
make the calculation more tractable, some of which are discussed in detail in references \cite{Koiller02A,Andres81}. First, the
Heitler-London expression for the exchange energy is approximated
by the Coulomb exchange integral alone, an approximation that is
asymptotically quite good, however it is this approximation that
is responsible for the difference between our result and that obtained
by KHD in reference \cite{Koiller02}. In reference \cite{Koiller02A}, the authors have used the complete form for the Heitler-London expression, and obtained results that match those presented here. 
\par The other approximation made is to effectively ignore the
contribution of the periodic part of the Bloch functions
$u_\mu({\bf r}) =1$; this is an excellent approximation as can be
seen from the figures in which it is almost impossible to
distinguish between the results obtained in this approximation and
those for which the detailed Bloch structure was included. The ability to ignore
the periodic part of the Bloch structure in the computation  of the Heitler-London integrals significantly reduces the complexity of the
calculation by eliminating the sums over reciprocal lattice vectors. It is
worth examining this approximation in more detail as we find it is
effectively the same approximation as is made in effective mass formalism to obtain the 
Kohn-Luttinger form for the donor electron wavefunction.
\par The Heitler-London integrals are all of the form
\begin{eqnarray}
I &=& \int \Psi^*({\bf r}_1 -{\bf R}/2) \Psi^*({\bf r}_2 + {\bf
R}/2) V({\bf r}_1,{\bf r}_2) \Psi({\bf r}_2 -{\bf R}/2) \Psi({\bf
r}_1+{\bf R}/2) d{\bf r}_1 d{\bf r}_2   \nonumber\\ &=&\int
\sum_{\alpha,\beta,\gamma,\delta} \sum_{i,j,l,m} F_i({\bf r}_1
-{\bf R}/2) F_j({\bf r}_2 +{\bf R}/2) F_l({\bf r}_2 -{\bf R}/2)
F_m({\bf r}_1 +{\bf R}/2) V({\bf r}_1,{\bf r}_2)\nonumber\\ &\times& {\rm e}^{i ({\bf
k}_i +{\bf k}_m -{\bf k}_j -{\bf k}_l).{\bf R}/2} {\rm e}^{i ({\bf
k}_m -{\bf k}_i).{\bf r}_1} {\rm e}^{i ({\bf k}_l -{\bf k}_j).{\bf
r}_2} \nonumber\\ &\times&A^*_{{\bf k}_i,{\bf G}_\alpha} A^*_{{\bf
k}_j,{\bf G}_\beta} A_{{\bf k}_l,{\bf G}_\gamma} A_{{\bf k}_m,{\bf
G}_\delta} {\rm e}^{i ({\bf G}_\alpha -{\bf G}_\delta).{\bf r}_1}
{\rm e}^{i ({\bf G}_\beta -{\bf G}_\gamma).{\bf r}_2}  d{\bf r}_1
d{\bf r}_2.
\label{eq:integralform}
\end{eqnarray}
The various potentials represented in the above expression as
$V({\bf r}_1,{\bf r}_2)$ are all impurity potentials of a similar nature to those
encountered in the effective mass formalism, and so 
we expect that they are sufficiently slowly varying that we can
safely  ignore the rapidly oscillating terms in the above
integrand, in exactly the same manner. This immediately gives
\begin{equation}
I = \int  \sum_{i,j} F_i({\bf r}_1 -{\bf R}/2) F_j({\bf r}_2 +{\bf
R}/2) F_i({\bf r}_2 -{\bf R}/2)  F_j({\bf r}_1 +{\bf R}/2) V({\bf r}_1,{\bf r}_2) {\rm
e}^{i ({\bf k}_i +{\bf k}_m -{\bf k}_j -{\bf k}_l).{\bf R}/2}
d{\bf r}_1 d{\bf r}_2,
\end{equation}
which is equivalent to setting $u_\mu({\bf r}) =1$ in Eq.~\ref{eq:integralform}.
The ``rapidly oscillating'' terms include the exponentials
containing the values of ${\bf k}$ at the conduction band minima, which are separated 
by a minimum magnitude of $\delta {\bf k} = 2\sqrt{2} \times 0.85 \pi/d$. This separation is sufficiently large to allow us to ignore terms except those for which $i=m,l=j$. Since $\sum_{\bf G}  A^*_{{\bf k},{\bf G}} A^*_{{\bf k},{\bf G}} =1$, this allows us to ignore the periodic part of the Bloch functions in the Heitler-London Integrals.
\section{Fabrication}
\label{section:fab}
The observed extreme sensitivity of the exchange coupling on the
relative orientation of the two phosphorus donors sets stringent
requirements on the placement of donors for any quantum computer
architecture that is reliant on the exchange interaction to couple
qubits. In this section we discuss these implications for the
construction of a Kane nuclear spin quantum computer.
\par
\begin{figure}
\rotatebox{90}{\resizebox{5cm}{!}{\includegraphics{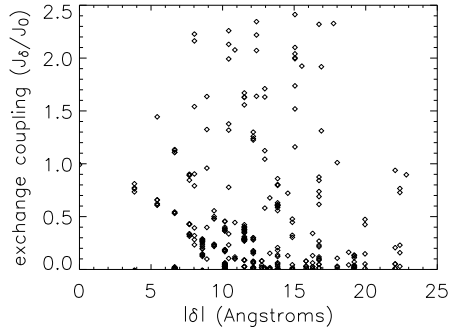}}}
\caption{\label{fig:displaced} Calculated exchange couplings for
donors at fcc lattice sites that are displaced by a vector ${\bf
\delta}$ from their ideal separation of 200.91\AA \ in the [100]
direction. The couplings are plotted as a fraction of the expected
exchange coupling $J(200.91$\AA$) =0.18 \mu$eV.}
\end{figure}

 Fabrication of the Kane solid-state quantum computer is being pursued
along two parallel paths \cite{Dzurak01}. In the so called
``bottom up'' approach, individual phosphorus donors are
effectively placed with atomic precision on a phosphorus surface by application 
of phosphine gas to a hydrogen terminated silicon surface in which individual 
hydrogen atoms have been removed at the proposed donor site using a scanning-tunnelling 
microscope tip. The hydrogen mono-layer is then removed and the surface overgrown with 
phosphorus \cite{Obrien01}. Although this approach has not been fully developed, it is likely that 
it will be able to produce an array of donors located at precise fcc substitutional sites,
 to within a few lattice spaces. Small displacements are inevitable however and can have
dramatic effects on the exchange coupling between the donors. This is illustrated in 
Fig.~\ref{fig:displaced} where we show a plot of the exchange coupling as a function of the
magnitude of the displacement $|\delta|$ of the second phosphorus dopant from its
desired position at a fcc substitutional site a distance of
200.91\AA, in the [100] direction, from the first dopant. These
dopants have been displaced by no more than four fundamental
lattice spacings on either of the two fcc lattices that make up
the silicon matrix, however the exchange coupling varies by more
than an order of magnitude emphasising the need for precise
placement.
\par
\begin{figure}
\rotatebox{90}{\resizebox{5cm}{!}{\includegraphics{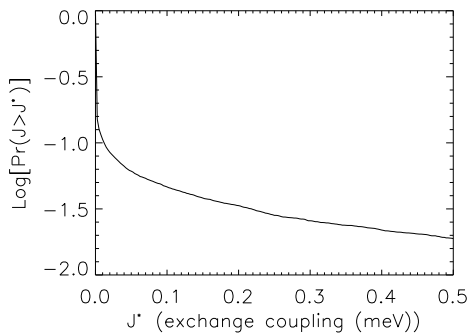}}}
\caption{\label{fig:probJ} A plot of the logarithm of the
integrated probability for the exchange coupling between donor electrons 
for donors implanted 200\AA \ apart
in the [100] direction, using a $15$keV phosphorus ion beam. The
donors are assumed to take a substitutional fcc lattice site.}
\end{figure}

 The second approach, known as the the top down approach calls for the implantation if the
phosphorus donors into the silicon using an ion beam of phosphorus ions incident on the substrate
\cite{Clark02}. The precision of such a technique in the placement
of dopants is fundamentally limited by the phenomenon known as ``straggling'',
whereby the implanted ions scatter from the nuclei of the host
silicon atoms. Simulations of this process for a beam of 15keV
phosphorus ions implanted into silicon, an energy appropriate for an implantation 
depth of approximately 200\AA \ into the silicon substrate, 
give a roughly Gaussian distribution for the final
position of the dopant with a variance of about 90\AA \ in the
transverse direction, and 100\AA \ in the longitudinal direction.
Using these data, for two dopants implanted 200\AA \ apart in the
[100] direction, we have calculated the distribution of exchange
couplings between the donor electrons, assuming that after thermal annealing  the phosphorus
donors take up a position on a fcc substitutional site.
Fig.~\ref{fig:probJ} shows a plot the integrated
probability, $\int_J^\infty P(J') dJ'$, that is the probability that
the exchange coupling greater than a value $J$, for
 two dopants implanted in the manner just described.
\par In the case of electron spin quantum computers where it is the
exchange energy that directly couples neighbouring qubits, a coupling of 
50$\mu$eV corresponds to a gate time of the order of $10$GHz. The top down approach
can in this case produce qubits coupled by extremely fast gates with a very high probability.
In nuclear spin quantum computers however the exchange interaction only mediates the inter-qubit coupling and the resulting gates are slower. The original Kane proposal calls for an  adiabatic implementation of a controlled-NOT gate that requires a exchange coupling of
approximately 50$\mu$eV\cite{Kane98,Fowler03}. Additional proposals exist for non-adiabatic 
implementations\cite{Wellard02a,Hill03} for
which controlled-NOT gates can be accurately implemented
with a lower exchange coupling. 
Fig.~\ref{fig:probJ} shows the probability that the bare 
exchange coupling for implanted donors being greater than  50$\mu$eV, is about $6 \%$. What is more important than the bare coupling however is the values of the exchange coupling that can be achieved with the application of J-gate biases.
\par We note that KHD have performed calculations that suggest a method
for overcoming the strong dependence of the the exchange coupling
on the donor orientation \cite{Koiller02A}. They calculate the
exchange coupling, in the same approximations as discussed
earlier, for phosphorus donor electrons in uni-axially strained
silicon. The strain is a product of the the silicon host being
fabricated on a layer of Si$_{1-\delta}$Ge$_\delta$, and is found
to suppress the oscillations in the coupling for donors that lie within a 
plane perpendicular to the direction of the
uni-axial strain. The coupling remains highly sensitive to
displacements away from this plane.
\section{External Control of the Exchange coupling}
\label{section:vdep} Inherent in the Kane proposal for a solid
state quantum computer is the ability to control the exchange
coupling of neighbouring donor electrons through the application
of voltage biases to a control J-gate placed on the substrate
surface between the position of the phosphorus donors, as
illustrated in Fig~\ref{fig:kanepic}. In this section we calculate the effect on 
the exchange coupling on the J-gate bias. Some results have been reported on similar
calculations. Fang {\it et al} \cite{Fang02} used an unrestricted
Hartree-Fock method, which avoids some of the approximations
inherent in the Heitler-London approach, to calculate the exchange
coupling between phosphorus donor electrons as a function of a
simplified electric potential. They use a trial wavefunction of
the form $\Psi({\bf r}) = F(r) \phi_{{\bf k}_\mu}({\bf r})$,
including only one of the six degenerate conduction band minima,
with a spherical envelope function, that is one of the same form
as Eq.~\ref{equation:envelope}, but with $a_\perp = a_\parallel =
20$\AA. As a result of their single minimum approximation they
cannot possibly reproduce the oscillatory nature of the coupling
which results from interference between the terms localised at the
six minima. They model the potential produced inside the device by the control gate as
one-dimensional potential of the form $V_J(x) = \mu
\frac{(x-R/2)(x+R/2)}{(R/2)^2} \times 30$meV, where $x$ is the
distance is along the direction between the two donors which are
situated at $x=\pm R/2$, as defined in Fig.~\ref{fig:kanepic}.
\par
\begin{figure}
\rotatebox{90}{\resizebox{5cm}{!}{\includegraphics{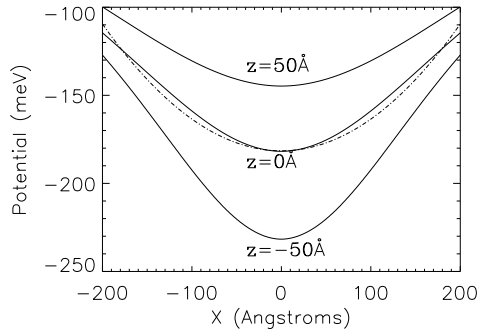}}}
\caption{\label{fig:potential} Potential produced inside the
device by a voltage bias of $1 V$ applied to a metallic J-gate.
Here $x$ is the inter-donor axis, the donors are located at $x=\pm
100$\AA \ and at $z=0$. The potential is independent of the
$y$-coordinate, mimicking a very long electrode. The dashed line
is the one-dimensional potential used by Fang {\it et al} with
$\mu = 0.15$ and shifted to match our potential at $x,z = 0$.}
\end{figure}

 In our calculation we compute the potential produced inside
the device due to the application of a voltage bias to a metallic
$J$-gate by numerical solution of the Poisson equation using a
commercial package,TCAD \cite{TCAD}, designed for the
semi-conductor industry. In Fig.~\ref{fig:potential} we plot the
potential parallel to the inter-donor axis, for several values of
the $z$ co-ordinate, which denotes the distance below the oxide
barrier, located at $z=-200 $\AA. For more information on the
details of the potential calculation see Ref~\cite{Pakes02}. We
compare the potential obtained for a $J$-gate voltage of $1$V, with the
one-dimensional potential of Fang {\it et al} and see that the two
potentials agree well for $z=0$, the plane of the donors, provided
we set $\mu = 0.15$. However we see that for this value of $\mu$
the two potentials are quite different in the planes 50\AA \ above
and below the donors. We must therefore conclude that the
one-dimensional approximation is not a good one.
\par To calculate the effect of the applied potential on the electron
we expand the wavefunction as follows:
\begin{equation}
\Psi({\bf r }-{\bf R}/2,V_J) = \sum_{n,l,m}  c_{n,l,m}(V_J) \sum_\mu
F^{n,l,m}_\mu({\bf r}-{\bf R}/2) {\rm e}^{i {\bf k}_\mu . ({\bf
r}-{\bf R}/2)},
\end{equation}
where the $F_\mu^{n,l,m}({\bf r}-{\bf R}/2)$ are non-isotropic
hydrogenic envelope functions defined by Faulkner
\cite{Faulkner69}. The $\mu$ is a label for the six degenerate
conduction band minima and determines the direction of the
anisotropy of the envelope function, for example $F^{n,l,m}_{\pm
z}({\bf r}) = F^{n,l,m}(x,y,\gamma z)$, is the hydrogenic function
with a Bohr radius $a = a_\perp$, and $\gamma =
a_\parallel/a_\perp$. The orthonormality of this basis is enforced
by the ${\rm e}^{i ({\bf k}_\mu - {\bf k}_\nu).({\bf r} - {\bf
R}/2)}$ terms which appear in the matrix elements and due to their
rapidly oscillating nature average to zero unless ${\bf
k}_\mu = {\bf k}_\nu$. The coefficients $c_{n,l,m}$ are determined
by direct numerical diagonalisation of the Hamiltonian in the
presence of the electrostatic potential. We use a basis of 140
states, which includes all states with principle quantum number up
to and including $n=7$, and we re-scale the energies such that the
unperturbed ground-state energy gives the experimentally observed
value of $45.5$meV for phosphorus donors in silicon. 
\par
\begin{figure}
\rotatebox{270}{\resizebox{5cm}{!}{\includegraphics{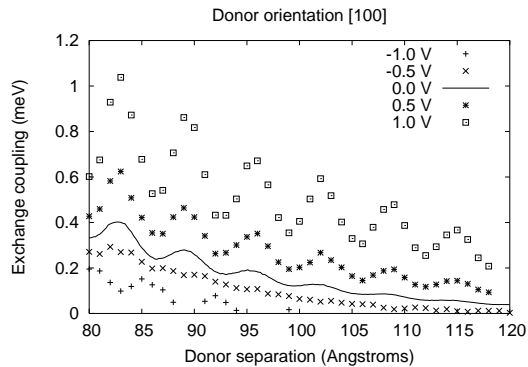}}}
\caption{\label{fig:vdep} Exchange coupling as a function of donor
separation for donors oriented along the [100] direction and various J-gate biases. }
\end{figure}

 The donor electron wavefunctions obtained in this way are then
used in the Heitler-London formalism to calculate the electron exchange energy,
 as a function of both the donor separation, and applied J-gate bias. 
The results are plotted in Fig.~\ref{fig:vdep}, for donor separations
from 80-120\AA\  in the [100] direction. We see that application of
a positive bias will tend to increase the
exchange coupling, while a negative bias decreases the strentgh of the coupling.
It is also worth noting that the relative increase in the exchange coupling produced by a
given  bias increases with the donor separation. Also the change in coupling is strongest at the peaks, the application of a positive bias enhances the oscillations in the exchange coupling whereas a negative bias of magnitude 1V inverts the oscillations such that points that were originally at peaks become troughs.
\par
\begin{figure}
\rotatebox{90}{\resizebox{5cm}{!}{\includegraphics{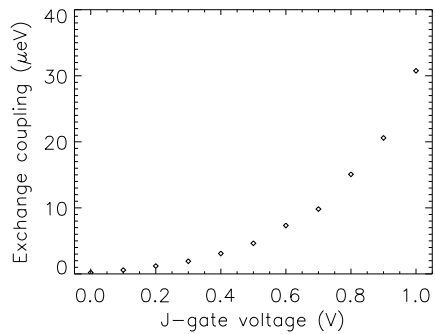}}}
\caption{\label{fig:vdep200} Exchange coupling as a function of
J-gate bias for donors at fcc lattice sites separated by 200.91\AA \ in the [100] direction.}
\end{figure}

 The range of
separations shown in these plots is far less than the
approximately 200\AA\ donor separation called for in the original
Kane proposal-we have plotted values over range for ease of
comparison with our previous results, and more particularly those
of KHD. In Fig.~\ref{fig:vdep200}  we plot the exchange
coupling as a function of the J-gate bias for donors separated by
200.91\AA\ in the [100] direction. At this separation
a 1V J-gate bias can increase the the exchange coupling by over
two orders of magnitude up to a value of approximately 30$\mu$eV. The exchange coupling can
also be reduced via application of a negative bias.
\par Although it is difficult to compare our results with those of
Fang {\it et al} due to the different potentials, the shift of an
order of magnitude obtained for separation of 200\AA\ and a bias
of 1V is greater than the shift they predict. For a value of $\mu
= 0.15$ Fang {\it et al} predict a shift of less than an order of
magnitude. This discrepancy can in part be attributed to the $1-D$
approximation for the gate potential which does not account for the fact that the
magnitude of the potential increases as the J-gate is approached
from below (in the $z$-direction). We note that it may not be
possible to apply voltages much greater than one volt to these
nano-scale devices without exceeding the breakdown fields of the
various materials involved \cite{Pakes02}.
\section{Discussion and conclusions}
In agreement with the approximations implicit in the calculations
of donor electron exchange energy of Koiller {\it et al}
\cite{Koiller02} we show explicitly that the periodic part of the
Bloch functions in the donor electron wavefunctions can be ignored
in the Heitler-London integrals, greatly reducing the complexity
of the calculation. This has allowed the calculation of the
intrinsic exchange coupling for a large number of donor pairs distributed
according to the probability distribution for donors implanted
200\AA \ apart by 15keV ion beams. In this manner we have determined
the probability distribution of the exchange coupling for the
top-down approach for the fabrication of a Kane solid-state
quantum computer.
\par In addition we have investigated the
application of control gate biases to increase the exchange
coupling between donor pairs using both realistic potentials and
realistic donor electron wavefunctions . We find that as expected
the exchange coupling can be increased by the application of a
positive bias, and decreased with a negative bias. 
Over the range of donor separations investigated it was found that the 
relative increase of the exchange coupling produced by a 
given bias increases with the separation, and for donors separated by 200\AA \ 
in the [100] direction a 1V bias can increase the coupling by over two orders of
magnitude.
\section{Acknowledgements} This work was supported by the
Australian Research Council and the Victorian Partnership for
Advanced Computing Expertise Grants scheme. The authors would like
to thank W. Haig (High Performance Computing System Support Group,
Department Of Defence) for computational support. HSG would like to
acknowledge the support of a Hewlett-Packard fellowship.

\widetext
\end{document}